\documentstyle [epsfig,preprint,aps,tighten]{revtex}
\draft
\begin{document}
\title{ A Novel SO(3) Picture for Quantum Searching}
\author{
 Gui Lu Long${}^{1,2,3}$, Chang Cun Tu${}^{1}$, 
Yan Song Li${}^{1}$, Wei Lin Zhang${}^{1}$ and Hai Yang Yan${}^{1}$ }
\address{
${}^{1}$Department of Physics, Tsinghua University, 
                Beijing 100084, China\\
${}^{2}$Institute of Theoretical Physics, Chinese Academy of Sciences, 
                Beijing 100080, China \\
${}^{3}$Center of Nuclear Theory, National Laboratory of Heavy Ion Accelerator, 
               Chinese Academy of Sciences, Lanzhou, 730000, China}
\maketitle
\begin{abstract}
An $SO(3)$ picture of the generalized Grover's quantum searching 
algorithm,with arbitrary unitary transformation and with arbitrary phase 
rotations, is constructed. In this picture, any quantum search operation
is a rotation in a 3 dimensional space. Exact formulas for the rotation angle 
and rotational axis are given. The probability of finding the marked
state is just $(z+1)/2$, where $z$ is the $z$-component of the state
vector. Exact formulas for this probability is easily obtained. 
The phase matching requirement and 
the failure of algorithm when phase mismatches are clearly explained. 
\end{abstract}

Grover's quantum search algorithm \cite{r1,r2} is one of the most 
celebrated quantum computing algorithms. It has been shown that 
the algorithm is optimal \cite{r3}. The algorithm can be generalized 
to arbitrary initial amplitude distribution \cite{r4}. It has many 
important applications, for instance, in the Simon problem \cite{r5} 
and quantum counting \cite{r6}. In the case where multiple marked state 
is involved, it can even search the data by just one query \cite{r6'}. 
Recently, it has been generalized to an arbitrarily entangled initial 
states \cite{r7}. Since Grover's algorithm involves only simple gate 
operations, it has been realized in 2 qubits \cite{r8,r9,r10},
 and 3 qubit NMR systems \cite{r11'}.

Grover's original algorithm has a simple geometric interpretation 
\cite{r2,r12,r14}.
When the Hadmard transformation is substituted by any 
arbitrary unitary transformation, it has been shown there is an $SU(2)$ 
group structure in the generalized algorithm \cite{r2,r14}. However, 
when generalizing the algorithm to arbitrary phase rotations, phase 
matching is vital \cite{r15,r16}. In \cite{r16} we have given an 
approximate formula for the amplitude of the marked state. But it is 
difficult to understand the phase matching requirement, as it is 
contrary to what one expects from an continuity argument.

In this Letter, we give a novel $SO(3)$ picture of the generalized 
quantum search algorithm by exploiting the relation between $SO(3)$ 
and $SU(2)$. In this $SO(3)$ picture the process of quantum search 
is crystalline transparent. The behavior of the algorithm with phase
matching or mismatching are clearly understood. 
This helps us to understand the various aspects of the algorithm, and to
further develop the algorithm.

The operator for quantum search\cite{r2} 
can be written as $Q=-I_\gamma U^{-1}
I_\tau U$, where $|\tau\rangle$ is the marked state, $|\gamma\rangle$ 
is the prepared state, usually $|\gamma\rangle$$=|0\rangle$. For 
arbitrary phase rotations, $I_\gamma =I-(-e^{i\theta}+1)$ $|\gamma
\rangle\langle\gamma|$, $I_\tau =I-(-e^{i\phi}+1)$ $|\tau\rangle
\langle\tau|$. In the basis where $
			|1\rangle=U^{-1}|\tau\rangle$,
			$|2\rangle=-(|\gamma\rangle-U_{\tau\gamma}
U^{-1}|\tau\rangle) 					/\sqrt{1-|
U_{\tau\gamma} |^{2}}$,
$Q$ can be written as 
\begin{equation}
	Q=\left(
		\begin{array}{cc}
	-e^{-i{\phi \over 2}}(\cos{\theta \over 2}+i\cos2\beta\sin{\theta
\over 2})
	    & -ie^{-i {\phi\over 2}}\sin2\beta\sin{\theta \over 2}	
	 \\
	-ie^{i {\phi\over 2}}\sin2\beta\sin{\theta \over 2}
	    & -e^{i{\phi \over 2}}(\cos{\theta \over 2}-i\cos2\beta\sin
{\theta \over 2})
		\end{array}
	\right),
\end{equation}
where we have written $U_{\tau\gamma}=e^{i \xi}\sin\beta$ (in Grover's 
original algorithm, $U_{\tau\gamma}={1\over\sqrt{N}}$, $\xi=0$, $\sin
\beta={1\over\sqrt{N}}$), and an overall phase factor has been neglected.

It is easy to check that $det(Q)=1$, and $Q$ is an element of the 
$SO(3)$ group. As is well known, each unitary matrix $u$ in $SU(2)$ 
group corresponds to a rotation $R_u$ in $SO(3)$ group \cite{r14'}. 
Here operator $Q$ corresponds to the rotation,
\begin{equation}
	\left( \begin{array}{ccc}
		R_{11} & R_{12} & R_{13} \\
		R_{21} & R_{22} & R_{23} \\
		R_{31} & R_{32} & R_{33} 
	\end{array}\right)
\end{equation}
where $		R_{11}=\cos\phi(\cos^2 2\beta\cos\theta+\sin^2 2\beta)
			+\cos2\beta\sin\theta\sin\phi$,
	$	R_{12}=\cos2\beta\cos\phi\sin\theta
			-\cos\theta \sin\phi	$,
$		R_{13}=-\cos\phi \sin4\beta \sin^2 {\theta \over 2}
			+\sin2\beta \sin\theta \sin\phi $,
$		R_{21}=-\cos(2\beta)\cos\phi\sin\theta
			+(\cos^2 {\theta \over 2}
			-\cos4\beta\sin^2 {\theta \over 2})\sin\phi$,
$		R_{22}=\cos\theta \cos\phi
			+\cos2\beta \sin\theta \sin\phi	$,
$		R_{23}=-\cos\phi \sin2\beta \sin\theta
			-\sin4\beta \sin^2 {\theta \over 2} \sin\phi $,
$		R_{31}=-\sin4\beta\sin^2 {\theta \over 2}	$,
$		R_{32}=\sin2\beta\sin\theta$,
$		R_{33}=\cos^2 2\beta+\cos\theta\sin^2 2\beta$.

A spinor in $SU(2)$ which describes the state of the quantum computer, 
$\Psi=\left(\begin{array}{c}a+bi\\c+di\end{array}\right)$ corresponds 
to a vector in $R^3$
\begin{equation}
	{\bf r}=\Psi^{\dag} {\bf \sigma} \Psi
		=\left(\begin{array}{c}
			x	\\
			y	\\
			z	\\
		\end{array}\right)
		=\left(\begin{array}{c}
			2(ac+bd)	\\
			2(-bc+ad)	\\
			a^2 +b^2 -c^2 -d^2\\
		\end{array}\right).
\end{equation}
The probability of finding the marked state is $P=a^2 +b^2 =(z+1)/2$. 
The $z$ component of the polarization vector is a measure of the probability.
For instance, the evenly distribution state
$\Psi_o=(\frac{1}{\sqrt{N}}$,
$\frac{\sqrt{N-1}}{\sqrt{N}} )^{\dag} $, 
corresponds to vector ${\bf r}_o =(2\sqrt{1- {1\over N}} \sqrt{1\over
N},
0,-1+ {2 \over N})^{T} $, which is nearly parallel to the $-z$ axis 
when $N$ is large. The marked state $\psi_a =(1,0)^{\dag}$,  corresponds
to ${\bf r}_a =(0,0,1)^T$, which is on the $+z$ axis. Thus the process 
of quantum searching in the $SO(3)$ picture is to rotate the state 
vector from a position nearly parallel to $-z$ axis to $+z$ axis.

The rotational
axis of (3) can be found by solving the eigen-value problem,
$	R_u {\bf l}={\bf l}$. This gives
$	{\bf l}=\left(\begin{array}{ccc}
			\cot{\phi \over 2} &
				1	&
	-\cot 2\beta \cot {\phi \over 2}+\cot{\theta \over 2}\csc 2\beta
	\end{array}\right)^{T}$.
Each iteration of $Q$ rotates about this axis an angle
\begin{equation}
	\alpha =\arccos[{1\over 4}(\cos4\beta+3)\cos\theta\cos\phi
		+\sin^2 2\beta({1\over 2}\cos\phi-\sin^2 {\theta\over 2})
		+\cos2\beta \sin\theta \sin\phi],
\end{equation} about the rotational axis.
In Grover's original algorithm, $\theta = \phi =\pi$, the rotation axis 
is exactly the $y-$axis, and the rotational angle is equal to the maximum
value of $4\beta$ (remember the relation between $SU(2)$ and $SO(3)$ , 
this corresponds an angle of $2\beta$ in the $SU(2)$). The state vector 
${\bf r}$ is being rotated within the $x-z$ plane from approximately $-z$
to $+z$ axis, where the marked state achieves maximum probability 
amplitude. The number of step requires to reach $+z$ axis is 
${{\pi-2\beta} \over \alpha}$$\approx 0.785\sqrt{N}-0.5$$\approx 
0.785\sqrt{N}$. The trace of tip of the state vector is shown in Fig.1.

In the most general case with arbitrary $\theta$ and $\phi$, the trace 
of the tip of state vector ${\bf r}$ is a circle. The state vector 
spans a cone with the top at the origin. During the rotation, the vector 
${\bf r}-{\bf r}_o$ is orthogonal to the rotational axis ${\bf l}$ at 
any time: $({\bf r}-{\bf r}_o)\cdot{\bf l}=0$. If  the 
state vector passes through $+z$ axis, 
that is ${\bf r}=(0,0,1)^T$ be in the trace, by solving equation ${\bf
r}
-{\bf r}_a \cdot {\bf l}=0$, we  have 
$	\cot{\phi \over 2} =\cot{\theta \over 2}$,
or $\phi=\theta$, the phase matching requirement which has been found 
in an approximate manner. However, the rotational axis is now
$	{\bf l}=\left(\begin{array}{ccc}
			\cos{\phi \over 2}	&
			\sin{\phi \over 2}	&
			\cos{\phi \over 2}\tan\beta
	\end{array}\right)^T$,
which is no longer the $y$  axis. The rotation angle is 
\begin{equation}
	\alpha=\arccos\{2[(\cos2\beta-1)\sin^2{\phi\over 2}+1]^2-1\}.
\end{equation}
If $N $ is very large, 
$
	{\bf l}\approx\left(\begin{array}{ccc}
			\cos{\phi \over 2}	&
			\sin{\phi \over 2}	&
			0
	\end{array}\right)$,
which is in the $x-y$ plane, and the initial state vector is nearly 
the $-z$  axis. The trace of tip of the state vector is a circle in 
the $x-z$ plane. Each interation rotates the state vector an angle 
$\alpha$ given by $(11)$. To first order in $\beta$, $\alpha\approx 
4\beta \sin\frac{\phi}{2}$, which corresponds a rotation of $2\beta
\sin\frac{\phi}{2}$ in $SU(2)$. The number of steps requires to seach 
the marked state is larger than that in the original version, as given 
in \cite{r16}. However in this case , the centre of the circle is no 
longer the origin. The state vector can pass the $+z$  axis, that is,it can 
reach the marked state with near certainty, but not the $-z$ where the 
amplitude of the marked state is zero. This has clearly been demonstrated
in the numerical calculation in Ref. \cite{r16}. 

When $\theta\neq\phi$, the trace the tip of the state vector is still a 
circle. But it is very tilted. In Figure 2, it is drawn for the case of 
$\theta={\pi\over 2}$, $\phi={\pi\over 10}$. Here we see the rotating axis
is nearly the $z$  axis, the circle span by the state vector tip is nearly
parallel to the $x-y$ plane.  Therefore the amplitude of the marked state
can not reach $1$, neither can it reach zero. This explains naturally
the intringuing narrowlly bounded behavor of the algorithm we have found
 in Ref. \cite{r15}.

To summarize, we have given a novel $SO(3)$ interpretation of the 
quantum search algorithm. In this picture , the effect of quantum 
search is clearly displayed. In particular, the phase-mismatching 
are clearly understood. This throws new light on the algorithm, and 
we hope it helpful for further development of the algorithm.

\begin{figure}
\begin{center}
\epsfig{figure=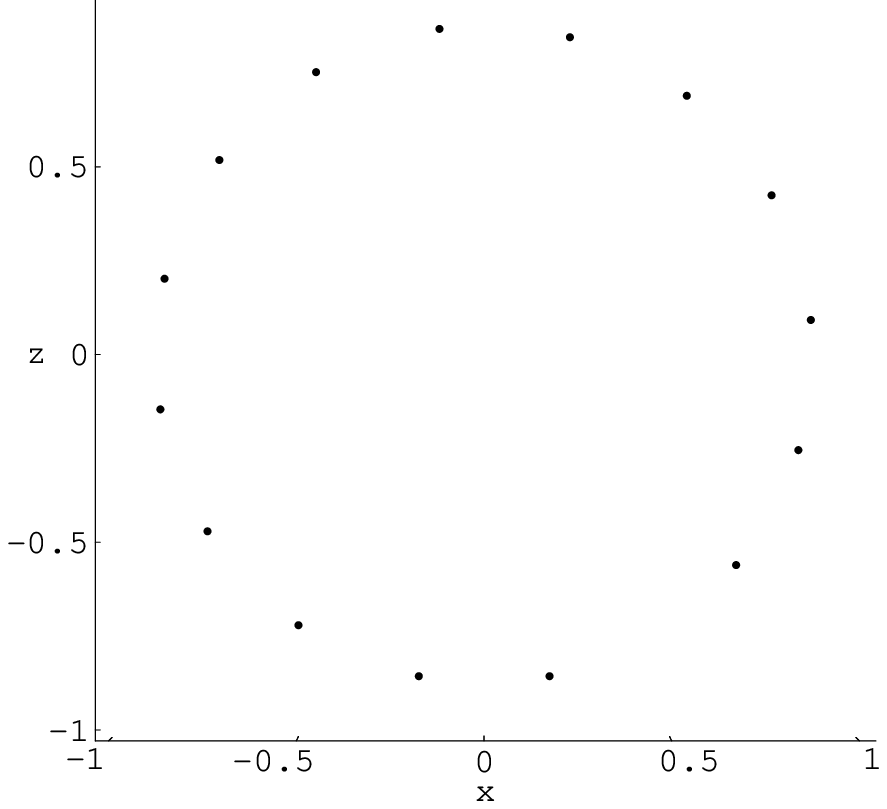,width=8cm}
\caption{The trace of the vector state tip when phase matching is
satisfied.} 
\end{center}
\end{figure}

\begin{figure}
\begin{center}
\epsfig{figure=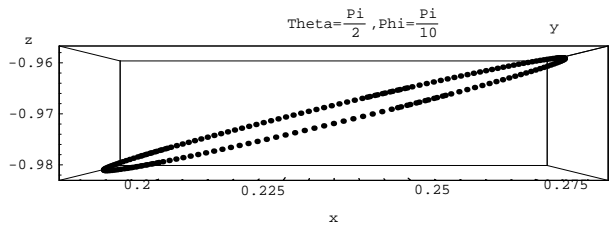,width=8cm}
\caption{3D plot of the trace of the vector state tip when phase mismatches.} 
\end{center}
\end{figure}
\end{document}